# Asymmetric transportation induced by thermal noise at the nanoscale

Rongzheng Wan[1], Jun Hu[1] and Haiping Fang[1,] [*]

[1]*Division of Interfacial Water and Laboratory of Physical Biology, Shanghai Institute of Applied Physics, Chinese Academy of Sciences, P.O. Box 800-204, Shanghai 201800, China*

[*]Corresponding author (email: fanghaiping@sinap.ac.cn)

**Based on a simple model, we theoretically show that asymmetric transportation is possible in nanoscale systems experiencing thermal noise without the presence of external fluctuations. The key to this theoretical advance is that the correlation lengths of the thermal fluctuations become significantly long for nanoscale systems. This differs from macroscopic systems in which the thermal noises are usually treated as white noise. Our observation does not violate the second law of thermodynamics, since at nanoscale, extra energy is required to keep the asymmetric structure against thermal fluctuations.**

Asymmetries in both structures and potentials are frequently exploited in nanoscale systems. In the past few decades, asymmetric transportation has been extensively observed in the systems that break the spatial inversion symmetry, particularly with nano- and biosystems [1-12].

The conventional ratchet theory, initiated from Smoluchowski in 1912 [13] and Feynman later [14], indicate that asymmetric transportation can exist only under the existence of an additional stochastic perturbation with long-range time correlation [15-21], while the thermal noise is regarded as white noise with negligible time correlation; however, in some nanoscale systems external fluctuations have not been actively applied [5, 8-9, 22-23]. Nanoscale systems usually display behavior different from bulk systems. At the nanoscale, the thermal noise may not be able to be treated as white noise. For example, at room temperature, the instantaneous velocity of an atom is at the level of ~100 m/s considering that the average kinetic energy in each degree of freedom is $k_BT/2$. In a liquid, the time duration between two collisions is on the order of 10 to 100 picoseconds. This is consistent with the statement by Magnasco that the timescale of the thermal fluctuations is smaller than ~100 ps [18, 24]. We note that the autocorrelation time of the thermal noise in bulk water at room temperature is at the order of 10 ps from molecular dynamics (MD) simulations which shows that the non-white behavior of the thermal noise should be taken into account at least in the MD simulations (see in the Appendix). The experiments and numerical simulations also show the unique behavior at the nanoscale. In 1997, Kelly *et al*. [22, 23] designed a molecule containing triptycene with helicenes, and observed spontaneous unidirectional rotations of the triptycene over a short time period. Takano *et al.* placed myosin on an actin filament and allowed myosin to move along the filament, and found that myosin exhibits a unidirectional Brownian motion along the filament [8].

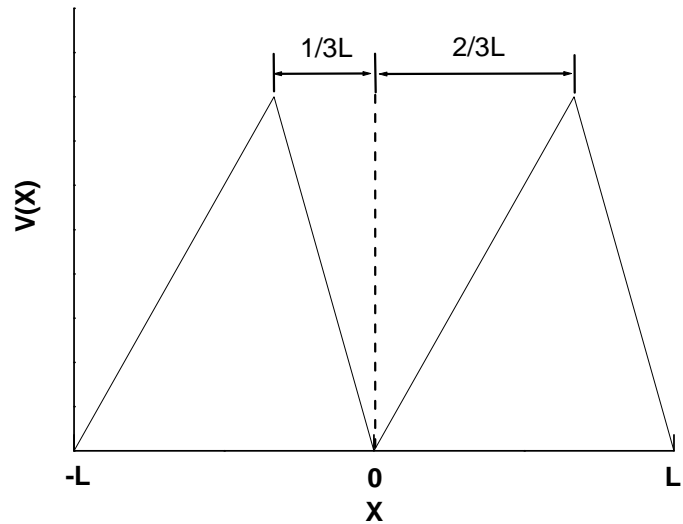

**Figure 1.** An example of the periodic asymmetric potential.

Here we show that the thermal noise may drive particles into unidirectional motion in nanoscale systems, based on the break from spatial inversion symmetry and the underlying physics if we consider the non-white behavior of the thermal noise in a nanoscale system. We note that, different from macroscopic systems, external energy is required to fix or constrain an item at the nanometer scale, since the length scale of the motion due to the thermal fluctuations is large enough to change the asymmetrical property of the systems. Thus, the system we considered is not an isolated system, i.e., extra energy from outside the system is required. Thus, our observation does not violate the second law of thermodynamics. Our finding may be possible to demonstrate universal kinematics at the nanoscale.

We use a simple model to illustrate the idea. The model system includes a particle in a periodical ratchet potential $V(x)$ (Fig. 1). The particle motion is governed by the Langevin equation of motion

$$m\ddot{x}(t) + V'(x(t)) = -\eta\dot{x} + \xi(t) + F(t) ,  \qquad (1)$$

where $m$ is the mass of the particle, $x$ is the position of the particle at time t, $\eta$ is the friction coefficient, and $F(t)$ is the external force. $\xi(t) = \sqrt{2\eta k_B T}\xi_0(t)$ is the thermal noise [18], where $k_B$ is the Boltzmann constant and $T$ is the temperature. $V(x)$ takes the form of

$$V(x+L_0) = V(x) = \begin{cases} \dfrac{3x}{2L_0}V_0 & 0 \leq x \leq \dfrac{2}{3}L_0 \\ \dfrac{3}{L_0}(L_0-x)V_0 & \dfrac{2}{3}L_0 < x < L_0 \end{cases}, \qquad (2)$$

where $L_0$ is the period of $V(x)$ and $V_0$ is the height of the potential barrier.

For the nanoscale model system we set $L_0 = 2$ nm, $V_0 = 2\ k_B T \approx 8.28\times10^{-20}$ J, and $m = 3.73\times10^{-26}$ kg, which is close to the mass of a water molecule. Different choices for the parameter values will not affect the conclusion obtained here as long as they remain at the nanoscale.

We introduce the following dimensionless units: the mass of the particle $m = 3.73\times10^{-26}$ kg, the length $x = 0.67$ nm, and the potential barrier $\Delta V = 2k_B T \approx 8.28\times10^{-21}$ J. Consequently, the time, force, and friction coefficient have units of 2 ps, 6.21 pN, and $1.86\times10^{-14}$ kg s$^{-1}$, respectively. Eq. (1) becomes

$$\ddot{\hat{x}}(\hat{t}) + \hat{V}'(\hat{x}(\hat{t})) = -\hat{\eta}\dot{\hat{x}} + \hat{\xi}(\hat{t}) + \hat{F}(\hat{t}) \tag{3}$$
$$\hat{\xi}(t) = \sqrt{2\hat{\eta}}\xi_0(t)$$

$$\hat{V}'(\hat{x}+3) = \hat{V}'(\hat{x}) = \begin{cases} 1 & 0 \le \hat{x} \le 2 \\ -2 & 2 < \hat{x} < 3 \end{cases} \tag{4}$$

We choose $\hat{\eta} = 0.1$ for the nanoscale system [25]. The simulation results shows that other choice of the values (i.e., $\hat{\eta} = 10$) do not affect the conclusion.

We consider the case without any external force, i.e., $\hat{F}(\hat{t}) = 0$. The noise portion is represented by $\hat{\xi}(t) = \sqrt{2\hat{\eta}}\xi_0(t)$ [21], where $\xi_0(t)$ satisfies [24]

$$\langle \xi_0(t) \rangle = 0$$

$$\langle \xi_0(t)\xi_0(t') \rangle = 0 \text{ for } |t-t'| \ge \tau_0,$$

and $\tau_0$ is the duration time of a collision. At the macroscale, $\tau_0$ is very small in the limit $\tau_0 \to 0$, resulting in $\langle \xi_0(t)\xi_0(t') \rangle = \delta(t-t')$ where $\xi_0(t)$ is white noise. Here, we assume that the duration time $\tau_i$ is a random number sample from a Gaussian distribution function with a mean value of $\Delta T$: $\langle \tau_i \rangle = \Delta T$; also, the noise during each $\tau_i$ has a sine distribution:

$$\hat{\xi}(t') = \sqrt{2\hat{\eta}}\xi_0 \sin(\pi t'/\tau_i), \quad 0 \le t' < \tau_i.$$

The autocorrelation time of this random fluctuation is approximately equal to $\Delta T$, and the time correlation function of this random fluctuation is quite similar to the time correlation function of the thermal noise in bulk water obtained by MD simulations (see in the Appendix). In the macroscopic and mesoscopic view, where the timescale is much greater than $\Delta T$, $\hat{\xi}(t')$ can be considered as white noise.

We performed simulations at different values of $\Delta T$ with a time step of $10^{-3}\hat{t}$ and $5\times10^8$ steps for each system.

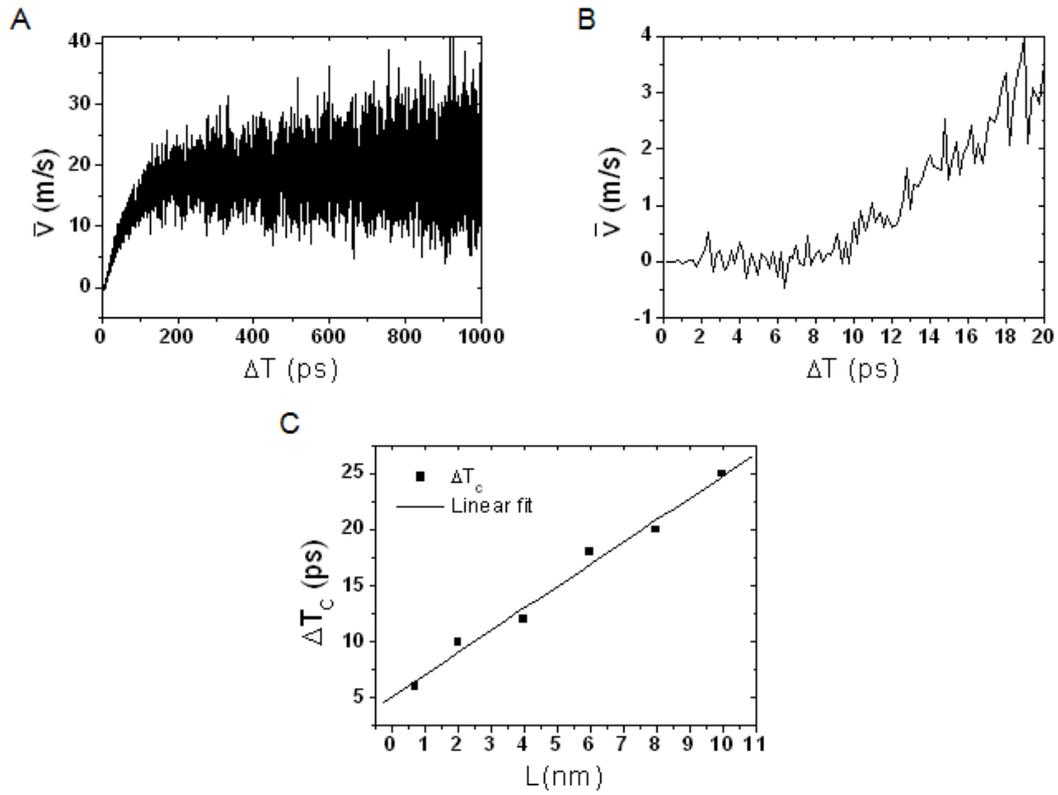

**Figure 2.** (**A**) Average velocity of the particle experiencing noise $\xi(t)$ with different average time scales $\Delta T$. (**B**) An enlargement showing $0 < \Delta T < 20$ ps. (**C**) Critical time period $\Delta T_C$ for different length period $L$ of the ratchet potential

Fig. 2(A) shows the average velocity $\bar{v}$ of the particle for different $\Delta T$. The average velocity is close to 0 until $\Delta T$ reaches a critical value, then $\bar{v}$ increases significantly as

$\Delta T$ increases, and becomes saturated at about 17 m/s. From Fig. 2(B) we observe that the critical value of the average time scale $\Delta T$ is around 10 ps. We define this critical time scale as $\Delta T_C$; it is the maximum value of $\Delta T_A$ such that, when $\Delta T < \Delta T_A$, the average of $\bar{v}$ within a neighborhood of 0.4 ps is below 0.1 m/s. These observations show that the biased particle motion in this system is possible only when the time scale of the noise is larger than a critical value of 10 picoseconds, for an asymmetrical system having a typical nanometer length scale. Fig. 2(C) shows the critical time period $\Delta T_C$ for different length period $L$ of the ratchet potential. From this figure we observe that $\Delta T_C$ decreases with L linearly in the range of 0.7 nm $\leq$ L $\leq$10 nm.

To further study the physics for such biased motion induced by random noise, we consider the case with only one component of the noise, i.e., $\xi(t) = \zeta(t) = A\sin(2\pi t/\Delta T)$, and keep the other parameters of the system unchanged.

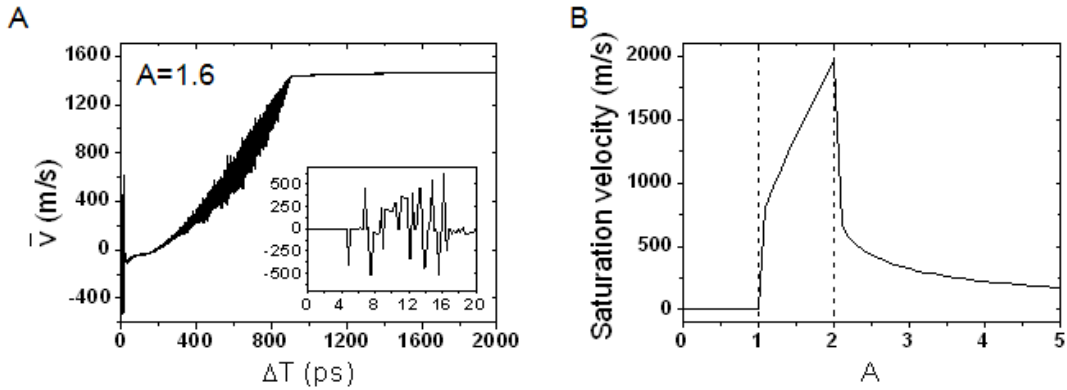

**Figure 3.** **(A)** Average velocity $\bar{v}$ of the particle under nanoscale perturbations of the form $\zeta(t) = A\sin(2\pi t/\Delta T)$, $A$=1.6 (inset shows an enlargement), **(B)** Saturation velocity with different A.

Fig. 3(A) shows the average velocity $\bar{v}$ for typical values of $A$. In the case of $A=1.6$, there is no clear biased motion of the particle for $\Delta T < 4.6$ ps. In the range 4.8 ps $\leq \Delta T \leq$ 20 ps, the particle moves both forward and backward; such current reversals have also been observed in a variety of systems at the meso- and macroscale under the influence of external fluctuations with long-range correlations [15-17]. $\bar{v}$ fluctuates about 0 until $\Delta T$ reaches a critical value of 180 ps, after which $\bar{v}$ increases with some fluctuation as $\Delta T$ increases to 900 ps; at that point the average velocity reaches saturation and remains constant for $\Delta T > 900$ ps. Comparing with Fig. 2, we can see that if $A$ becomes stochastic, the critical time scale $\Delta T_C$ decreases to 10 ps. This is much less than the critical value of 180 ps required for sine perturbation which have a certain $A$ and $\tau_i$. Further simulations show that the fluctuations with a stochastic $\tau_i$ and constant $A$ require a $\Delta T_C$ of 90 ps. These suggest that the $\Delta T_C$ in the case of the stochastic fluctuation (thermal noise, *etc.*) is much lower than the $\Delta T_C$ in the case of the fluctuation with a certain strength and time period. Fig. 3(B) shows the saturation velocity with respect to $A$. When $A \leq 1$, the saturation velocity of the particle is always very close to 0. The saturation velocity increases sharply for $1 < A \leq 2$, and then decreases for $A > 2$.

Next we consider the mesoscale system and use the following example to illustrate the idea. We assume $L_0 = 2$ μm, $V_0 = 8.28 \times 10^{-15}$ J, and $m = 3.73 \times 10^{-20}$ kg, and for convenience we introduce dimensionless units as with the nanoscale system to further transform these quantities to $\tilde{m} = 10^{-6} \times \hat{m}$, $\tilde{L}_0 = 10^{-3} \times \hat{L}_0$, and $\tilde{V}_0 = 10^{-6} \times \hat{V}_0$. The governing equation now has the same form as Eqs. (3) and (4).

The thermal noise has an average strength of $k_BT$, which corresponds to the value of $A = 0.0014$ in the dimensionless quantity in the mesoscale. From Fig. 3(B), it is clear that there is no biased motion induced. Even with a very low probability, some fluctuation reaches the range of $A > 1$, and biased motion only happens when $\Delta T > 180$ ns (See Fig. 3(A) but the time scale should be multiplied by 1,000). Thus, the average velocity of the particle is negligibly small in mesoscale or macroscale systems, and asymmetric transportation is negligibly small with only thermal noise even though the system has asymmetry both in its structure and potential.

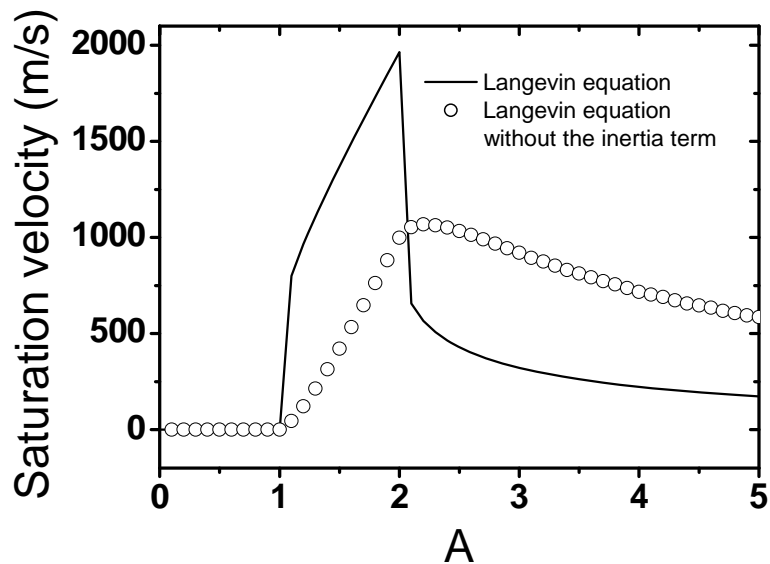

**Figure 4.** Saturation velocities for different values of A from the Langevin equation (solid line) and for the case where the inertia term of $m\ddot{x}(t)$ is dropped (circles).

The Fokker-Planck equation is usually used in conventional ratchet theory with overdamped systems, and when this equation is derived from the Langevin equation the inertia term $m\ddot{x}(t)$ is ignored. This approximation may cause large errors in the nanoscale system with low viscosity. Fig. 4 shows the saturation velocities obtained from the Langevin equation and those when the inertia term $m\ddot{x}(t)$ is dropped, for the nanoscale system described above. A clear difference can be seen. Interestingly, the critical values of $A$ above which the saturation velocity becomes clearly positive are consistent in both simulation methods.

To summarize, counter to the conventional view, we show that a particle in a nanoscale system can be driven by the unbiased fluctuations with the autocorrelation time around 10 ps. This suggests that the biased motion of particles due to thermal noise may be possible in nanoscale systems if spatial inversion asymmetry holds considering that the time scale for the thermal noise is on the scale of 10 to 100 picoseconds. Interestingly, the autocorrelation time of the thermal noise in bulk water obtained from the traditional MD simulations is at the order of 10 ps (see in the Appendix), comparable to the critical time scale $\Delta T_C$ for the nanoscale system with noise. We note that this does not violate the second law of thermodynamics, since the thermal noise also impacts the asymmetrical structure of the system, i.e., without the external energy to hold the asymmetrical structure, the unidirectional motion only exists in a very short time period [22, 23]. This external energy makes the nanoscale system a non-equilibrium system, so that the detailed balance and time reversibility is not required to be held in this system [26]. This is different from macroscopic systems in which we can fix an item without any extra energy, since the motion of the item due to thermal fluctuations is negligible.

There may be beneficial applications of this observation in biosystems in which spatial asymmetry occurs frequently. For example, the length of the Escherichia coli glycerol uptake facilitator (Glpf) is 4.8 nm [12], which is comparable to the length used in the model here. In the simple model used in the present manuscript, we used an auto-correlated noise in the model system which induced the unidirectional motion. For real systems, especially biosystems, the real thermal noise may be more complex, and the detail and effect of the thermal noise at nanoscale worth further studying. Moreover, there are more interesting phenomena and physics to be exploited in systems between the nanoscale and macroscale, where the thermal fluctuations may become weak both in terms of their influence on holding the asymmetry as well as their impact on particles moving in the asymmetrical system.

We gratefully acknowledge Profs. Bailin Hao, Jingdon Bao and Qing Ji for their helpful discussions. This work was supported by National Natural Science Foundation of China (10825520, 11175230), Shanghai Leading Academic Discipline Project (B111), the Knowledge Innovation Program of the Chinese Academy of Sciences, and Shanghai Supercomputer Center of China.

**Appendix: Time correlation profile of the thermal noise and random fluctuation**

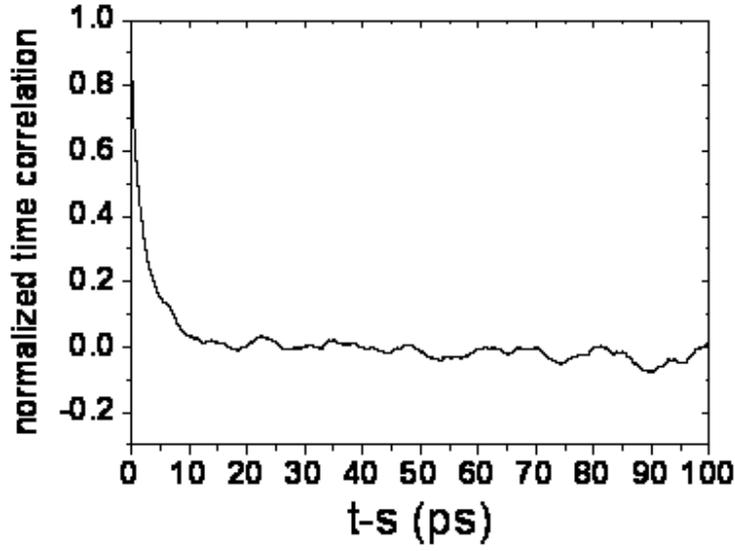

Figure A1. Normalized time correlation $\frac{\langle \xi(t)\xi(s) \rangle}{\langle \xi(t)\xi(t) \rangle}$ profile of the thermal noise in bulk water at room temperature (300K) calculated by MD simulations.

Fig. A1 shows the time correlation $\frac{\langle \xi(t)\xi(s) \rangle}{\langle \xi(t)\xi(t) \rangle}$ profile of the thermal noise in bulk water. Here, $\xi(t)$ is the total interaction force on an Oxygen atom from surrounding water molecules at time $t$ obtained from molecular dynamics (MD) simulation. The MD simulations were carried out at a constant pressure (1 bar with initial box size $L_x$=2.8 nm, $L_y$=2.8 nm, $L_z$=4.0 nm) using the coupling scheme of Berendsen [27] and temperature (300K) using the coupling scheme of Nosé-Hoover [28, 29] with Gromacs 3.3 [30]. The TIP3P water model [31] was applied, and a time step of 2 fs was used. Periodic boundary conditions were applied in all directions. From this figure, we can find that the autocorrelation time of the thermal noise in bulk water at room temperature from MD simulations is at a scale of 10 ps.

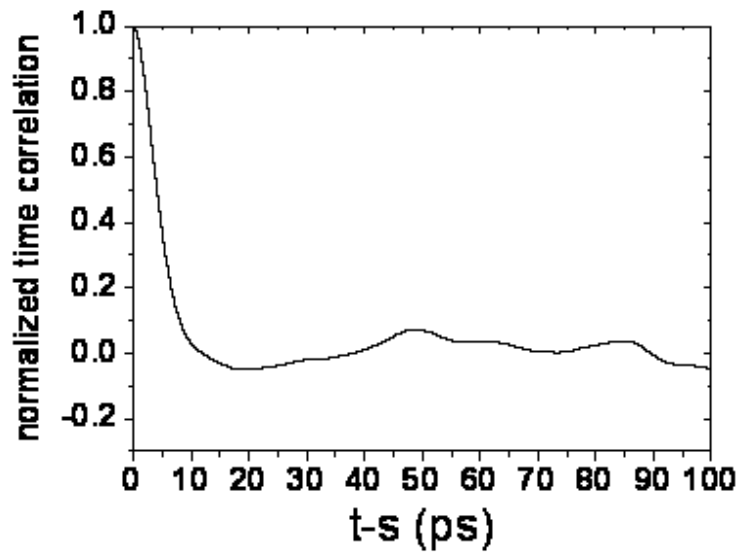

Fig. A2 Normalized time correlation profile of the random fluctuation we used in the manuscript with $\Delta T = 10$ ps.

Fig. A2 shows the normalized time correlation of the random fluctuation we used in our manuscript with $\Delta T = 10$ ps. From the figure we can see that this random fluctuation has an autocorrelation time around 10 ps. Moreover, by comparing this correlation function with the correlation function of the thermal noise in bulk water at room temperature (300K) calculated by MD simulations shown in Fig. A1, we can see that they have quite similar structures with comparable autocorrelation time at the nanoscale.